# Quantum Transport in 40-nm MOSFETs at Deep-Cryogenic Temperatures

T.-Y. Yang, A. Ruffino, J. Michniewicz, Y. Peng, E. Charbon, and M. F. Gonzalez-Zalba

*Abstract*—In this letter, we characterize the electrical properties of commercial bulk 40-nm MOSFETs at room and deep cryogenic temperatures, with a focus on quantum information processing (QIP) applications. At 50 mK, the devices operate as classical FETs or quantum dot devices when either a high or low drain bias is applied, respectively. The operation in classical regime shows improved transconductance and subthreshold slope with respect to 300 K. In the quantum regime, all measured devices show Coulomb blockade. This is explained by the formation of quantum dots in the channel, for which a model is proposed. The variability in parameters, important for quantum computing scaling, is also quantified. Our results show that bulk 40-nm node MOSFETs can be readily used for the co-integration of cryo-CMOS classical-quantum circuits at deep cryogenic temperatures and that the variability approaches the uniformity requirements to enable shared control.

## I. INTRODUCTION

SILICON-BASED electronics has shown great potential as a platform for quantum information technology [1]. From a physics perspective, single-electron spins can be confined in gate-defined quantum dots (QDs) to realize qubits [2-4], which can provide long coherence times and be operated close to fault-tolerance fidelity levels [5, 6], and elevated temperatures [7, 8]. From a technological perspective, QDs can be manufactured in a similar fashion to field-effect transistors (FETs) [9, 10] with a small footprint (100×100 $nm^2$). This gives the opportunity to leverage very large scale integration (VLSI) techniques to scale up the technology beyond state-of-the-art 2-qubit processors [6, 11-13] to two-dimensional QD arrays [14, 15], a requirement for fault-tolerant quantum computing [16].

However, so far silicon qubit systems require either (1) high precision e-beam lithography (EBL) for gate patterning on a planar Si or SiGe substrate [17], or (2) narrow etching of a thin nanowire in a fully depleted silicon-on-insulator (FD-SOI) substrate [18-20], which are not yet industry-ready for massive scale integrated circuit production. Hence, exploring the potential of existing industry-standard complementary metal–oxide–semiconductor (CMOS) technologies for QIP could be an efficient way to integrate Si quantum electronics into VLSI. Unlike previously proposed quantum platforms, focusing on FD-SOI custom devices [21, 22] or just CMOS-compatible FD-SOI nanowire gate-all-around transistors [9], this approach involves fully industry-standard bulk CMOS transistors, thus implying the benefits of the mass-production readiness.

Moreover, scalability of QD array architectures requires compact classical readout and control electronics to be operated in close proximity to the quantum processor. Using standard CMOS technology allows the co-integration of silicon quantum devices with classical analog and digital electronic circuits to reduce the overhead of qubit control and readout [23, 24].

In this letter, we investigate commercial bulk 40-nm MOSFETs as a platform for QIP. We present a statistical characterization of charge transport at room and deep-cryogenic temperatures on a number of devices sufficient for a proof of concept. We observe the formation of QDs in the channel at 50 mK and propose a model for how this occurs. Furthermore, we quantify the intra-die variability across different identical devices in parameters important for quantum computing scaling, such as the locations of the dots, the voltage to load the first measurable electron, relevant to shared control quantum computing architectures [25], and the gate lever arm, important for dispersive readout schemes [26, 27].

## II. EXPERIMENT

We study bulk MOSFETs in a commercial 40-nm process. The devices under test (DUT) are planar *n*-type low threshold voltage FETs with gate length $L_g$=40 nm and gate width $W_g$=120 nm. We characterize charge transport at 300 K and 50 mK in a dilution refrigerator (Oxford Instruments Triton). The drain voltage $V_{ds}$, gate voltage $V_g$, and drain current $I_d$ are applied and measured by a parameter analyzer (HP 4156A). We extract classical parameters such as threshold voltage $V_{th}$, subthreshold slope $SS$, transconductance $g_m$, and drain induced barrier lowering (DIBL) from a standard *I-V* measurement. We measure the Coulomb blockade parameters such as charging energy $E_c$, lever arm $\alpha$ and source-drain capacitance ratio $C_s/C_d$ from a stability map (Coulomb diamonds) at low drain bias. We present the data from 18 DUTs in total, which have nominally identical physical dimensions and we perform a statistical analysis.

This research has received funding from the European Union's Horizon 2020 research and innovation programme under grant agreement No. 688539 (http://mosquito.eu). M.F.G.Z. acknowledges support from the Royal Society.

T.-Y. Yang and M. F. Gonzalez-Zalba are with Hitachi Cambridge Laboratory, Hitachi Europe Ltd. Cambridge CB3 0HE, UK.

A. Ruffino, Y. Peng, and E. Charbon are with the School of Engineering, École Polytechnique Fédérale de Lausanne (EPFL), Lausanne, Switzerland.

J. Michniewicz is with University of Cambridge, Cambridge CB3 0HE, UK.

## III. RESULT AND DISCUSSION

We characterize the room temperature charge transport in DUTs by measuring $I_d$ as a function of $V_g$ ranging from -0.2 to 1.0 V with $V_{ds}$ ranging from 0.05 to 1.0 V, as shown in Fig. 1(a). The transport characteristics of $V_{th}$, $SS$, and $g_m$ can be extracted from the experimental data in Fig. 1 and are benchmarked in TABLE I. The ON/OFF ratio of DUTs is approximately $10^6$, and $SS\approx 86$ mV/dec is slightly higher than the theoretical limit. DIBL is expected in such short channel transistors and indeed it is observed and has a value of 162±27.85 mV/V at 300 K.

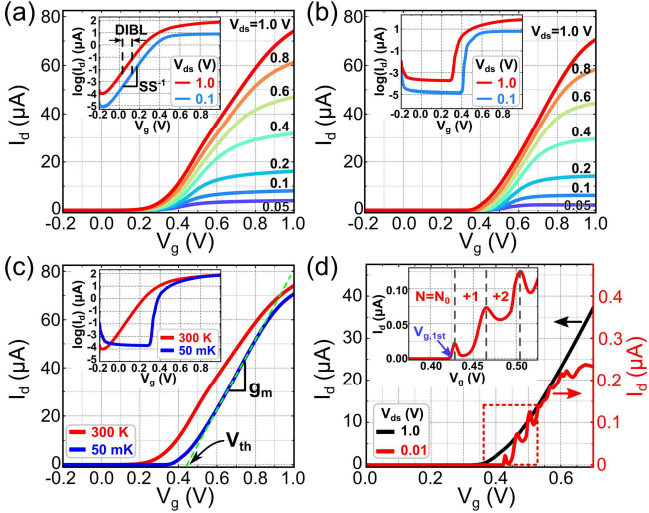

Fig. 1. Drain current $I_d$ vs $V_g$ at various $V_{ds}$ (a) at 300 K and (b) at 50 mK. Inset: logarithm scale. (c) $I_d$ vs $V_g$ at 300 K and 50 mK with $V_{ds}$=1.0 V. Inset: logarithm scale. The green dashed line is a linear fit. (d) $I_d$ vs $V_g$ at 50 mK. Inset: zoom-in of the red dotted region. N: number of electrons.

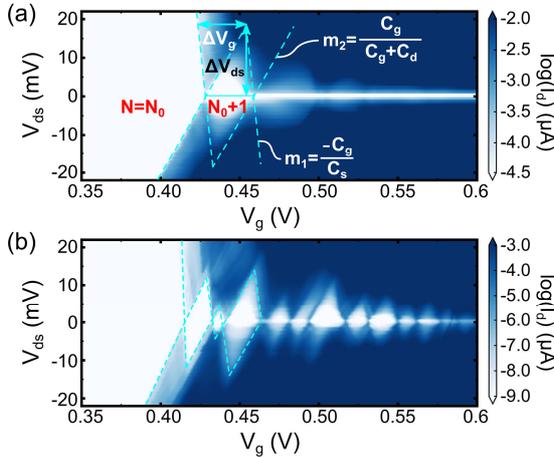

Fig. 2. Charge stability diagrams as a function of $V_g$ and $V_{ds}$ of devices (a) T01 and (b) T07. The measurement is carried out at 50 mK.

At 50 mK, with $V_{ds}$=0.1 V (Figs. 1(b)-(c)), we observe an increase in threshold voltage $\Delta V_{th}(=V_{th,50mK}-V_{th,300K})$=0.126 V, which can be understood by the higher Fermi potential at cryogenic temperatures [28]. $g_m$ increases more than 37% and $SS$ decreases to 9.93 mV/dec. Although both results indicate better power-efficiency at cryogenic temperatures, $SS$ is still far from the Boltzmann limit. Our data is in line with previous reports, where it saturates to a value proportional to the extent of the conduction band tail [29]. When we apply low $V_{ds}$,

Coulomb blockade oscillations are observed near $V_{th}$, as shown in Fig. 1(d). The presence of quasi-periodic Coulomb blockade peaks suggests the existence of a quantum dot (QD) in the channel. In low $V_{ds}$ and low $V_g$ regime, the current is dominated by the tunneling current through *Source-QD-Drain* instead of through the conventional inversion layer. Similar transport behaviours were previously observed in a p-type MOSFET platform as well [22]. The number of electrons (N) loaded into QD can be counted according to the appearance of Coulomb peaks, so that one is able to constrain the QD in the first measureable electron regime, i.e., $N=N_0+1$ where $N_0$ is the offset charge. It is worth mentioning that Coulomb blockade oscillations are observed in all DUTs in this paper. However, we notice that not only devices with quasi-periodic Coulomb diamonds (Fig. 2(a)) are observed, but also devices with irregular Coulomb diamonds (Fig. 2(b)), suggesting a double or multiple QD system.

TABLE I
COMPARISON OF PERFORMANCE AT $V_{ds}$=+0.1 AND +1.0 V AT 300 K AND 50 mK, AND QUANTUM DOT CIRCUIT PARAMETERS AT 50 mK.

| Temperature | 300 K | | 50 mK | |
|---|---|---|---|---|
| $V_{ds}$ (V) | 0.1 | 1.0 | 0.1 | 1.0 |
| $V_{th}$ (V) | 0.388±0.030 | 0.442±0.036 | 0.514±0.038 | 0.510±0.037 |
| SS (mV/dec) | 86.41±2.43 | 86.88±1.69 | 9.93±4.32 | 15.79±6.27 |
| $g_m$ (µS) | 21.48±1.79 | 118.07±7.48 | 29.51±6.02 | 144.53±10.48 |
| $\Delta V_{th}$ (V) | | | 0.126±0.017 | 0.067±0.011 |
| DIBL (mV/V) | 162.22±27.85 | | 113.33±24.67 | |
| $V_{g,1st}$ (V) | | | 0.488±0.039 | |
| α (eV/V) | | | 0.588±0.111 | |
| $C_s/C_d$ | | | 1.265±0.754 | |

For the single QD case, the full set of capacitances, the gate, source and drain capacitances, can be extracted from a Coulomb diamond diagram as in Fig. 2(a). The charging energy is $E_c=\frac{e^2}{C_\Sigma}=e\cdot\Delta V_{ds}$, where $C_\Sigma$ is the total capacitance to the quantum dot, namely $C_\Sigma=C_g+C_s+C_d$. We find $E_c$=18.4 meV, indicating that Coulomb oscillations should still be observable up to liquid helium temperature as $E_c>>k_BT$ at 4.2 K. This has also been confirmed by measurement. The capacitance between the gate and QD can be expressed as $C_g=\frac{e}{\Delta V_g}$, where $\Delta V_g$ is the gate voltage separation between adjacent Coulomb peaks. We find $C_g$=5.34 aF, corresponding to a QD with an equivalent diameter of 14.7±0.7 nm[1], which suggests that by reducing the channel width, for instance using 28-nm node, the probability of forming multiple QDs can be lowered. The capacitances between QD/Source ($C_s$) and QD/Drain ($C_d$) can be estimated from the slopes of the Coulomb diamond's boundaries as indicated in Fig. 2(a): $C_s=\frac{-C_g}{m_1}$ and $C_d=\frac{C_g(1-m_2)}{m_2}$. Thus,

---

[1] We estimate the size of the quantum dot by using a parallel disc model: $C_g=\varepsilon_r\varepsilon_0\pi r^2/L$, where $\varepsilon_r$ and $\varepsilon_0$ are SiO$_2$ permittivity and vacuum permittivity, respectively, $r$ is the radius of quantum dot disc and the separation between the gate electrode and quantum dot disc (EOT=) $L$=1 -1.2 nm is used based on *ITRS 2001* and *ITRS 2003*.

$$\frac{C_s}{C_d} = \frac{-m_2}{m_1(1-m_2)}. \quad (1)$$

And finally, we define the gate lever arm

$$\alpha = \frac{\Delta V_{ds}}{\Delta V_g} = \frac{C_g}{C_\Sigma}. \quad (2)$$

For multi-QD systems, as in Fig. 2(b), the measured $I_d$ is a result of multiple parallel paths: (i) *Source-QD1-Drain*; (ii) *Source-QD2-Drain*; (iii) *Source-QD1-QD2-Drain*, (iv) *Source-QD2-QD1-Drain* as in Fig. 4(a), and hence it is not possible to extract the full set of capacitances. However, by restricting our analysis to the first Coulomb oscillation, we obtain a statistical characterization of $V_{g,1st}$, the voltage where the first Coulomb oscillation occurs (Fig. 1(d)), $\alpha$ and the $C_s/C_d$ ratio across the die (Fig. 3). We find a fairly consistent $V_{g,1st}$=0.488±0.039 V, suggesting a small variation from DUT to DUT. A small gate control voltage variation, $\Delta V_g < \Delta V_{g,1st}$, is essential for quantum computing requiring shared control schemes [25]. We find that the DUTs approach this requirement, but further variability reduction is still necessary.

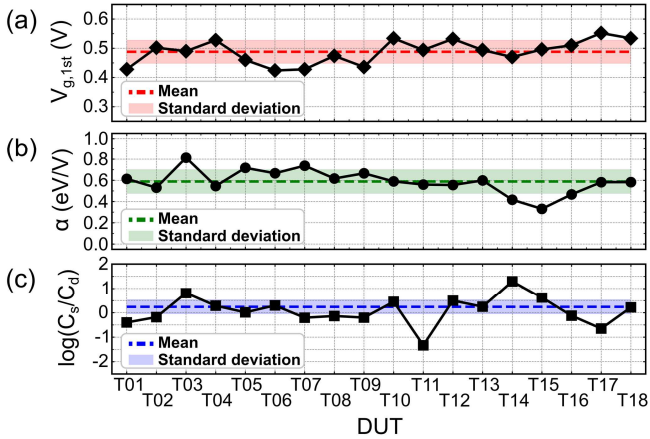

Fig. 3. Summary of (a) $V_{g,1st}$, (b) $\alpha$, and (c) $C_s/C_d$ of all DUTs in the paper. The dashed lines are mean values, and the shaded areas indicate (mean value)±(standard deviation).

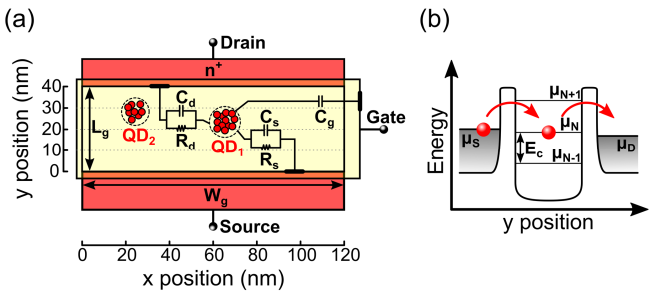

Fig. 4. (a) Schematic of a MOSFET top view. Red spheres represent charge carriers. (b) Schematic of energy vs channel position in y-axis in (a). $\mu_s$, $\mu_d$, and $\mu_N$ are the electrochemical potentials of *Source*, *Drain*, and *QD*, respectively. When $V_{ds}(=\mu_s-\mu_d)$ is small, $I_d$ only occurs when $\mu_s$, $\mu_d$, and $\mu_N$ are in alignment.

For the lever arm, we find $\alpha$=0.6±0.1 eV/V (Fig. 3(b)), a value comparably larger than other planar quantum dot devices [30-32] and just below those reported for 3D geometries [33]. Large $\alpha$ is essential in dispersive readout schemes [18, 26, 27, 33] and 40-nm bulk MOSFET should provide a good platform. Finally, the $C_s/C_d$ ratio can be used for estimating the location of QD in the channel, since in general the capacitance is inversely proportional to the separation between two conductors. $C_s/C_d$ of all DUTs is summarized in Fig. 3(c), and the result suggests that the QDs tend to locate well centered in the channel with good reproducibility and an average deviation of 0.2±6.5 nm from the center. Therefore, we conclude that the QDs are formed from the charge carrier accumulation due to the applied gate bias rather than from the dopants close to *Source* and/or *Drain* diffused during the implantation and activation annealing processes. Also, we rule out the *p*-type dopants in the body of the *n*-MOSFET as the origin of Coulomb blockade since our measurements are performed in the voltage region close to the conduction band edge. Surface roughness and remote charges in the gate stack may as well contribute to the formation of the dots [34-35]. With all these considerations in mind we depict a to-scale schematic of how Coulomb blockade transport occurs in 40-nm bulk MOSFETs, see Fig. 4.

IV. CONCLUSION

This paper presents a statistical characterization of 18 commercial 40-nm MOSFETs at room and deep-cryogenic temperatures (50 mK). DUTs behave as classical MOSFETs at 300 K and at 50 mK under high bias, with improved performance. At 50 mK under low bias, observed Coulomb oscillations indicate that QD systems are formed in the channel in the subthreshold region. We have statistically characterized the properties of the QD systems, such as $V_{g,1st}$, gate coupling parameter $\alpha$, and dot-to-electrode capacitances $C_g$, $C_s$ and $C_d$, and found that these devices could be a useful resource for large-scale QIP given their low variability, planar geometry and high $\alpha$. Our results suggest that 40-nm MOSFETs can be used to build both classical circuits and quantum circuits or to co-integrate the two into quantum-classical hybrids at liquid helium temperatures and below.